\documentstyle[preprint,pra,aps]{revtex}
\begin{document}
\draft\onecolumn
\baselineskip=24pt

\title{\bf  Collective spontaneous emission in a 
q-deformed Dicke model}

\author{Stefano Mancini$^{\dag}$, Vladimir I. Man'ko$^{\ddag}$ 
and Paolo Tombesi$^{\dag}$}

\address{$\dag$ Dipartimento di Matematica e Fisica and 
Unit\`a INFM,\\ 
Universit\`a di Camerino, 
I-62032 Camerino, Italy\\
$\ddag$ Lebedev Physical Institute, Leninsky Pr. 53, 
117924 Moscow, Russia}

\date{Received: \today}

\maketitle\widetext

\begin{abstract}

The q-deformation of a single quantized radiation mode interacting 
with a collection of two level atoms is introduced, analysing its 
effects on the cooperative behavior of the system.

\end{abstract}


\widetext

\section{Introduction}

Since the fundamental paper by Dicke~\cite{Dicke} on collective 
spontaneous emission, a considerable amount of increasing interest 
has been devoted to an assembly of $N$ two-level atoms located within 
a distance much smaller than the radiation wavelength. Such a system 
has commonly been referred to as the Dicke model, in which the 
cooperative 
nature of the spontaneous emission is assumed to result from indirect 
atom-atom coupling via the field mode only.

Since the erliest numerical investigations~\cite{numerical} the model 
has 
been known to present two limit cases when it may be treated as an 
almost linear system. These are called ``strong'' and ``weak'' field 
regimes 
and have been extensively studied  developing several approximation 
methods~\cite{approx}.

It was also shown~\cite{losses} that 
by considering only a fraction $s$ of the $N$ atoms inverted, it is 
possible to neglect the losses provide to have $s\ll N.$
In such a case, the model becomes exactly solvable, up to date, for 
$s\le 8$~\cite{Koz95}. 

On the other hand, the recent development of quantum 
groups~\cite{qgroups} 
has motivated great interest in q-deformed algebraic structures, 
in particular
the q-oscillators~\cite{qosc}. In this framework, there have been 
already 
examples to treat solvable model, like the Jaynes--Cummings, which 
permits
the application of quantum algebra~\cite{qJC}.

Here, we shall consider another physically important model of the 
interaction 
of a radiation field (q-deformed) with $N$ two-level atoms 
placed in a lossless single mode cavity, when only a part $s$ of 
the $N$ atoms radiates 
spontaneously in the presence of $N-s$ unexcited atoms, and we 
shall show how 
the deformation affects the collective phenomena.

In physical terms, this generalization allows us to introduce an 
additional parameter $q$ into the Dicke model, which shows, 
therefore, 
an intensity dependent coupling resembling that one of the 
Jaynes--Cummings~\cite{nlJC}, where only one atom is considered.

The physical nature of the additional parameter $q$ might be 
treated as 
q-nonlinearity characteristics of the q-oscillator vibrations, 
for which the frequency of vibrations specifically 
depends on the amplitude of the vibrations~\cite{Naples}.
In this case, the nonlinearity of the field vibrations results 
the 
intensity-dependent coupling with atoms in a cavity. On the 
other hand, the
intensity-dependent coupling could arise due to a highly 
nonlinear response
of the atom to the action of the linear vibrating-radiation-field
oscillator.  The use of q-deformed Dicke model, in this case,
corresponds to a
phenomenological description of the nonlinear interaction of 
particular atoms
with standard electromagnetic field. 

Thus, the aim of our work is to study how the nonlinearity 
described by the
q-oscillators influences the known properties of spontaneous 
emission of atoms
in a cavity in the framework of the Dicke model.

\section{The model}

The Hamiltonian of the system we wish to start with, in rotating 
wave 
approximation, can be written as \cite{Koz92} ($\hbar=1$) 
\begin{eqnarray}
H&=&H_0+V\,;\nonumber\\
H_0&=&\omega_f a^{\dag}a+\omega\sum_{j=1}^N S^{(3)}_j\,;\\
V&=&g\sum_{j=1}^N\left[a^{\dag}S^-_j
+aS^+_j\right]=g\left[a^{\dag}S^-+aS^+\right]\,\nonumber ,
\end{eqnarray}
where $a^{\dag}$ ($a$) is the photon creation (annihiliation) 
operator and $S^-_j$, $S^+_j$, and $S^{(3)}_j$ are the pseudo-spin 
lowering, rising and inversion  operators of the $j$th atom, 
respectively;
$\omega_f$ denotes the frequency of the field mode while $\omega$ 
is the atomic transition frequency. In what follows we assume 
exact resonance and choose the scale in such a way that 
$\omega_f=\omega=1$. Furthermore, within the 
small-sample approximation, the coupling coefficient $g$ 
is the same for all the atoms. 
The basis vectors of the model read
\begin{equation}
|s,m\rangle=|s-m\rangle_a\otimes|m\rangle_f\,,
\end{equation}
where $|m\rangle_f$ denotes the Fock state of the field, 
while $|s-m\rangle_a$ is the normalized symmetric Dicke 
state of the atomic subsystem with $s-m$ atoms excited~\cite{Koz92}. 
The initial condition will be $|s,0\rangle$
when $s$ atoms are initially excited and no photon is
present (spontaneous emission).

In the cases, $s=1$ and $s=2$, the problem is characterized 
by equidistant eigenvalues spectra~\cite{s12}, while the 
latter becomes already unequidistant for $s=3$, and the 
eigenvalues are incommensurate quantities~\cite{s3}. 
This unequidistance leads to the phenomenon of collective 
collapses and revivals of the system oscillations~\cite{Koz92}.

Let us now introduce the deformation of the field as~\cite{Naples}
\begin{eqnarray}
V_q&=&g\left[V^+_q+V^-_q\right]\,;\nonumber\\
V^+_q&=&a_qS^+=af(n)S^+;\\
f(n)&=&f(a^{\dag}a)=\sqrt{\frac{\sinh\,(n\log q)}{n\sinh\,(\log
q)}}\,,\nonumber
\end{eqnarray}
with $V_q^-$ the hermitian conjugate of $V_q^+$;
then the nonvanishing matrix elements of these operators are
\begin{eqnarray}
\langle s,m+1|V^-_q|s,m\rangle&=&
f(m+1)\sqrt{(m+1)(s-m)(N-s+m+1)}\nonumber\\
&=&\langle s,m|V^+_q|s,m+1\rangle\,.
\end{eqnarray}

\section{The time evolution}

In what follows, for simplicity, we restrict our analysis to the 
lower three values of $s$.
Let us first consider the case of $s=1$, then the system ket state 
can be written as 
\begin{equation}
|\Phi(t)\rangle=C_0(t)|1,1\rangle
+C_1(t)|1,0\rangle\,,
\end{equation}
with the initial condition $C_0(0)=0;\;C_1(0)=1.$
With this choice, the time-dependent Schr\"odinger equation (in 
the interaction picture) leads to the coupled equations
\begin{eqnarray}
i{\dot C}_0&=&gf(1)\sqrt{N}C_1\,;\\
i{\dot C}_1&=&gf(1)\sqrt{N}C_0\,.
\end{eqnarray}
Since $f(1)=1,\,\forall q$, the field deformation does not affect 
the oscillatory solutions.
In the case of $s=2$, we have
\begin{equation}\label{state}
|\Phi(t)\rangle=C_0(t)|2,2\rangle
+C_1(t)|2,1\rangle+C_2(t)|2,0\rangle\,,
\end{equation}
with equations
\begin{eqnarray}
i{\dot C}_0&=&gf(2)\sqrt{2N}C_1\,;\\
i{\dot C}_1&=&gf(2)\sqrt{2N}C_0+gf(1)\sqrt{2(N-1)}C_2\,;\\
i{\dot C}_2&=&gf(1)\sqrt{2(N-1)}C_1\,,
\end{eqnarray}
whose solutions are
\begin{eqnarray}
C_0(t)&=&\frac{2gf(2)\sqrt{N(N-1)}}{\Omega^2}
\left[\cos\,(\Omega t)-1\right]\,;\\
C_1(t)&=&\frac{-i\sqrt{2(N-1)}}{\Omega}
\sin\,(\Omega t)\,;\\
C_2(t)&=&\frac{2g(N-1)}{\Omega^2}
\left[\cos\,(\Omega t)-1\right]+1\,,
\end{eqnarray}
with
\begin{equation}\label{Om}
\Omega=\sqrt{2f^2(2)N+2(N-1)}\,.
\end{equation}

Now, the expectation value of the collective operator
$S^{(3)}=\sum S^{(3)}_j$, in the eigenstate $|s,m\rangle$ is
\begin{equation}
\langle s,m|S^{(3)}|s,m\rangle=-\frac {N}{2}+s-m\,,
\end{equation}
and represents the inversion of the atomic energy for the 
whole system of $N$ atoms;
but in turn we may use the operator $S^z=S^{(3)}+(N/2)-(s/2)$ 
obtaining
\begin{equation}
\langle s,m|S^z|s,m\rangle=\frac {s}{2}-m\,,
\end{equation}
which represents the inversion of the atomic energy for 
the group of $s$ atoms \cite{Koz92}. Its time evolution can 
be written as
\begin{equation}\label{Eat}
E^s_{\rm {at}}(t)=\langle\Phi(t)|S^z|\Phi(t)\rangle\,,
\end{equation}
and for the case of $s=2$ we get
\begin{equation}\label{Eat1}
E^{s=2}_{\rm {at}}(t)=C_2^2(t)-C_0^2(t)\,.
\end{equation}
It becomes clear from Eq.~(\ref{Om}) that the q-nonlinear 
deformation 
changes (reduces) the period of oscillations of the 
quantity~(\ref{Eat1})
making faster the beating of the energy between the 
atoms and the radiation
field in the cavity. 
Furthermore, the amplitude of these oscillations 
is reduced towards the upper values as can be seen in Fig.~1; 
this means that, due to the deformation, there do not exist times 
at which all initially excited atoms release the photons.
In the limit of extremely high deformation, the period of 
oscillations tends to zero and the value of~(\ref{Eat1}) 
tends to remain constant (equal to $+s/2$), which implies that 
for strong
nonlinearity the beating phenomenon disappears.

For $s=3,$ the analogous of Eq. (\ref{state}) takes the form
\begin{equation}
|\Phi(t)\rangle=C_0(t)|3,3\rangle
+C_1(t)|3,2\rangle+C_2(t)|3,1\rangle
+C_3(t)|3,0\rangle\,,
\end{equation}
and the equations for the coefficients become
\begin{eqnarray}
i{\dot C}_0&=&g f(3)\sqrt{3N}C_1\,;\\
i{\dot C}_1&=&g f(3)\sqrt{3N}C_0+2gf(2)\sqrt{N-1}C_2\,;\\
i{\dot C}_2&=&2g f(2)\sqrt{N-1}C_1+g\sqrt{3(N-2)}C_3\,;\\
i{\dot C}_3&=&g\sqrt{3(N-2)}C_2\,.
\end{eqnarray}
Their solutions can be written as
\begin{eqnarray}
C_0(t)&=&\frac{-2if(2)\sqrt{N-1}}{3gf(3)\sqrt{N(N-2)}}
\frac{\Omega^2_+\Omega^2_-}{\Omega^2_+-\Omega^2_-}
\left[\frac{\sin\,(\Omega_+t)}{\Omega_+}
-\frac{\sin\,(\Omega_-t)}{\Omega_-}\right]\,;\\
C_1(t)&=&\frac{2f(2)\sqrt{N-1}}{3Ng^2f^2(3)\sqrt{3(N-2)}}
\frac{\Omega^2_+\Omega^2_-}{\Omega^2_+-\Omega^2_-}
\left[\cos\,(\Omega_+t)
-\cos\,(\Omega_-t)\right]\,;\\
C_2(t)&=&\frac{-i\left(\Omega^2_+-3Ng^2f^2(3)\right)
\Omega_+\Omega_-^2}{3Ng^3f^2(3)\sqrt{3(N-2)}
\left(\Omega^2_+-\Omega^2_-\right)}\sin\,(\Omega_+t)\nonumber\\
&+&\frac{i\left(\Omega^2_--3Ng^2f^2(3)\right)
\Omega_+^2\Omega_-}{3Ng^3f^2(3)\sqrt{3(N-2)}
\left(\Omega^2_+-\Omega^2_-\right)}\sin\,(\Omega_-t)\,;\\
C_3(t)&=&\frac{\left(\Omega^2_+-3Ng^2f^2(3)\right)
\Omega_-^2}{3Ng^2f^2(3)
\left(\Omega^2_+-\Omega^2_-\right)}\cos\,(\Omega_+t)\nonumber\\
&-&\frac{\left(\Omega^2_--3Ng^2f^2(3)\right)
\Omega_+^2}{3Ng^3f^2(3)
\left(\Omega^2_+-\Omega^2_-\right)}\cos\,(\Omega_-t)\,,
\end{eqnarray}
where
\begin{eqnarray}
\Omega_{\pm}&=&\frac{g}{\sqrt{2}}\Bigg\{
\left(3f^2(3)+4f^2(2)+3\right)N-4f^2(2)-6\nonumber\\
&\pm&\Bigg[\Big(
9f^4(3)+24f^2(2)f^2(3)-18f^2(3)+16f^4(2)+24f^2(2)+9
\Big)N^2\nonumber\\
&+&\Big(36f^3(3)-32f^4(2)-24f^2(2)f^2(3)-72f^2(2)-36
\Big)N\nonumber\\
&+&\Big(16f^4(2)+48f^2(2)+36
\Big)\Bigg]^{1/2}\Bigg\}^{1/2}\,,
\end{eqnarray}
In this case, the inversion of the atomic energy of 
Eq.~(\ref{Eat1})
is given by
\begin{equation}
E_{\rm {at}}^{s=3}(t)=
\frac{3}{2}-3|C_0(t)|^2-2C_1^2(t)-|C_2(t)|^2\,,
\end{equation}
and its behavior is shown in Fig.~2, for various values of 
the deformation parameter. Substantially, the system oscillates 
with two incommensurate frequencies, then the phenomenon of 
quantum beats leads to modulated oscillations. However, collapses 
and revivals occur with different period if one introduces the 
field deformation, and they tend to disappear as soon 
as one increases the strength of the deformation.

We have used a relatively small value of $N$ in the figures 
for graphic convenience.
It is worth noting that by increasing the value of $N$ only 
the time scale of the described
effects change.

\section{Conclusions}

In conclusion, we have shown that the introduction of an 
algeabric q-deformation in the radiation field interacting 
with a collection of two-level atoms may lead to interesting 
effects such as the inhibition of the
collective  spontaneous emission. Really, a large deformation 
drastically
enhances the probability of induced emission/absorption, hence 
already with very few 
photons the atoms become saturated and the inversion  (of the 
group of $s$ atoms)
approaches $s/2$; from this point of view spontaneous emission 
is not suppressed, but the
probability of reabsorption of spontaneously emitted photons is 
strongly enhanced.

The analysis has been performed by considering the radiation field 
initially in a vacuum state, but one could study other cases as 
well.

Of course the results of this (oversimplifed) model are valid  
whenever
the cavity is engineered to substain exactly only one radiation 
mode \cite{dem},
otherwise the field deformation could lead to the decay over 
other modes. 

It is worth noting that in our work the character of the 
nonlinearity of the
deformed Dicke model is described by the specific function 
$f(n)$ in Eq.~(3).
Other types of possible nonlinearities were discussed in the 
framework of
nonlinear-oscillator models, for which q-oscillators were 
replaced by
f-oscillators with the same structure of annihilation operator 
but with a
different function $f(n)$ \cite{sudref}.   
As one can see from the solutions to the Schr\"odinger equation 
given by
(6)--(7), (12)--(14), and (25)--(28), the character of the used 
nonlinearity
does not influence the possibility to solve exactly the 
Schr\"odinger 
equation for the deformed Dicke model, at least for 
$s=1,\,2,\,3.$
 
Thus, our formulas are valid not only for q-oscillators but 
for generic
f-oscillators of the radiation field interacting with two-level 
atoms in the
cavity (in the framework of Dicke model) as well. For any 
nonlinearity, for
which the function $f(n)$ takes very large values, the quantum 
beats will be
suppressed as it takes place for the case of the discussed 
q-oscillator model. 
On the other hand, the concrete choice of the nonlinearity 
coded in the 
function $f(n)$ may demonstrate essentially different behavior 
of the collapses
and revivals from the ones given by the standard and q-deformed 
Dicke models. 
   
\section*{Acknowledgements}
We acknowledge the financial support of the Istituto Nazionale 
di Fisica Nucleare.

\newpage

FIGURE CAPTIONS

Fig. 1 Time evolution of the atomic inversion for 
$N=6$ and $s=2$. The deformation parameter takes the values:
a) $q=1$; b) $q=5$; c) $q=20$.

Fig. 2 Time evolution of the atomic inversion for 
$N=6$ and $s=3$. The deformation parameter takes the values:
a) $q=1$; b) $q=2$; c) $q=4$.

\end{document}